\documentclass[11pt,a4paper]{article}
\usepackage{cite}
\usepackage{amsmath, amsthm, amssymb,slashed}
\usepackage{dsfont}
\usepackage{eufrak}
\usepackage{yfonts}
\usepackage{float}
\usepackage[caption = false]{subfig}
\usepackage[final]{graphicx}
\usepackage{hyperref}
\hypersetup{
    colorlinks=false,
    linkcolor=blue,
    filecolor=magenta,      
    urlcolor=magenta,
}
 
\usepackage{graphicx}

\usepackage{fancyhdr}
\usepackage{mathbbol}
\usepackage{mathabx}
\usepackage{url}
\usepackage{color}
\usepackage{bbm}
\usepackage{dsfont}
\usepackage{bm}
\usepackage{amsfonts}

\usepackage{datetime}

\def\be{\begin{equation}}
\def\ee{\end{equation}}
\def\bea{\begin{eqnarray}}
\def\eea{\end{eqnarray}}

\def\bpm{\begin{pmatrix}}
\def\epm{\end{pmatrix}}
\parskip=\baselineskip

\def\1{{\bf 1}}
\def\2{{\bf 2}}
\def\3{{\bf 3}}
\def\4{{\bf 4}}

\def\hat{\widehat}
\font\teneurm=eurm10 \font\seveneurm=eurm7 \font\fiveeurm=eurm5
\newfam\eurmfam
\textfont\eurmfam=\teneurm \scriptfont\eurmfam=\seveneurm
\scriptscriptfont\eurmfam=\fiveeurm

\font\teneusm=eusm10 \font\seveneusm=eusm7 \font\fiveeusm=eusm5
\newfam\eusmfam
\textfont\eusmfam=\teneusm \scriptfont\eusmfam=\seveneusm
\scriptscriptfont\eusmfam=\fiveeusm

\font\tencmmib=cmmib10 \skewchar\tencmmib='177
\font\sevencmmib=cmmib7 \skewchar\sevencmmib='177
\font\fivecmmib=cmmib5 \skewchar\fivecmmib='177
\newfam\cmmibfam
\textfont\cmmibfam=\tencmmib \scriptfont\cmmibfam=\sevencmmib
\scriptscriptfont\cmmibfam=\fivecmmib

\begin{document}
\begin{titlepage}
\begin{flushright}

\end{flushright}
\vskip 1.5in
\begin{center}
{\bf\Large{A note on polarized light from Magnetars}}
\vskip 0.4cm {L.M. Capparelli$^*$,  A. Damiano$^*$, L. Maiani$^\P$ and  A.D. Polosa$^*$} \vskip 0.05in {\small{ \textit{$^*$Dipartimento di Fisica and INFN, Sapienza Universit\`a di Roma, \\ Piazzale Aldo Moro 2, I-00185 Roma, Italy}\vskip -.4cm
{\small{ \textit{$^\P$Theory Department, CERN,  Geneva, Switzerland.}}}
}
}
\end{center}
\begin{center}
\makeatletter
\tiny\@date
\makeatother
\end{center}
\begin{abstract}
In a recent paper it is claimed that vacuum birefringence has been experimentally observed for the first time by 
measuring the degree of polarization of visible light from a Magnetar candidate, a neutron star with a magnetic field presumably as large as $B\sim 10^{13}$~G.  
The role of such a strong magnetic field is twofold. First, the surface of the star emits, at each point, polarized light with linear polarization correlated 
with the orientation of the magnetic field. Depending on the relative orientation of the magnetic axis of the star with the direction to the distant observer, a certain degree of polarization should be visible. Second, the strong magnetic field in the vacuum surrounding the star could enhance the effective  degree of polarization observed: vacuum birefringence.  We compare experimental data and theoretical expectations concluding that the conditions to  support a claim of {\it strong} evidence of vacuum birefringence effects are not met.
\newline
\newline
PACS: 12.20.-m, 97.60.Jd, 14.80.Va 
\end{abstract}
%
\end{titlepage}

{\bf \emph{Introduction.}} In a recent paper~\cite{Mignani:2016fwz} the results of the observation of a  Magnetar in the constellation of Corona Australis are reported, 
showing an interesting indication of  linear polarization of light --- for the moment a $\approx 3\sigma$ effect, to be confirmed in forthcoming measurements.

Magnetars\cite{Duncan:1992hi} are stars with extremely intense magnetic fields, $B\sim 10^{12}\div 10^{14}$~G, as deduced from the study of their X-ray spectra. In the specific case of 
the Magnetar candidate analyzed (the isolated neutron star RX J1865.5-3754), the  emitted light appears  to follow a blackbody distribution indicating a surface temperature of $T\approx 10^6$~K.  The star radius is expected to be $R_{_{\rm NS}}\approx 10$~Km. 

According to a vast astrophysics literature, see {\it e.g.}~\cite{Meszaros}, the light emitted by the surface of a star with such a large magnetic field, should be polarized, with a definite (orthogonal) polarization with respect to  the direction of the magnetic field at every point of the star surface. However, even if each point on the star were to be considered as a 100\% linearly polarized light source, 
the distant observer (the Magnetar candidate discussed is estimated to be at a distance of 400 Ly)  will only see the superposition of the different sources and this results in a way smaller net polarization. Indeed, considering the small radius of the star, the orientation of the magnetic field on its surface varies  sensibly from point to point whereas the wave-vectors $\bm k$ are all parallel and directed to the observer: each emitted photon has a polarization which is simultaneously in a plane orthogonal to $\bm k$ and to $\bm B$, with the direction of $\bm B$ varying from point to point.  

Depending on the orientation of the magnetic axis of the star with respect to the observation line, different degrees of net polarization could be estimated.  We might observe here that there is a geometric upper bound to the observable degree of linear polarization, which is found in the case in which the magnetic axis and the observation line are orthogonal to each other (corresponding to the best possible observation conditions with the equatorial line of the star seen as a diameter of the star disk). Following for example a study by Pavlov and Zavlin, see Fig 4,5 and 7 in~\cite{Pavlov:1999cw}, it is clear that large degrees of polarizations can be observed in favorable observation conditions. We add that if the rotation axis of the star is significantly different from the magnetic one, one might expect that the geometrical polarization averages to smaller effective values considering the star rotation. However, in~\cite{Mignani:2016fwz, Ho:2007gs}  it is understood $\xi \lesssim 6^\circ$ for this angle, thus no significant averaging is expected.

In addition to this there is the possibility of an enhancement of the net   polarization observed as a consequence of  Quantum Electrodynamics (QED)  in presence of very strong magnetic fields $B\gtrsim B_{_{\rm QED}}=m^2/e\sim 10^{13}$~G. In this case, the Euler-Heisenberg interaction term is not negligible and its effect is that of providing a dielectric tensor $\epsilon_{ij}$ and a magnetic permeability tensor $\mu_{ij}$, as if the vacuum surrounding the star were a birefringent crystal. Thus, electromagnetic waves  in the neighbourhood of the star propagate with different refractive indices depending on whether $\bm E$ is parallel or orthogonal to the external $\bm B$ field. The difference $\Delta n$ between the refractive indices prevents the mixing of perpendicular and parallel modes, as an effective energy gap between the two would do. 

\begin{figure}[htb!]
\centering
   \includegraphics[width=7.5cm]{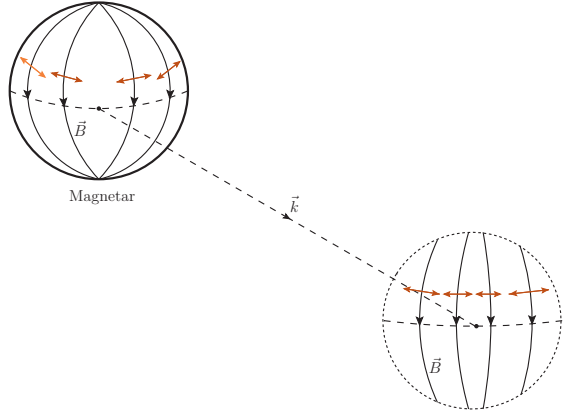}
 \caption{\small It is assumed that on the Magnetar, light is always polarized in a plane orthogonal to the magnetic field at a given point, and to the wave vector $\bm k$.  All distinct sources are seen as superimposed by the distant observer, resulting in a faint polarization.
Far away from the star, as long as  $\bm B$ is still appreciable, the curvature  of the surface tangent to the magnetic field vectors (spherical cap) is smaller: magnetic field vectors within the spherical cap  result approximately parallel to each other. However, if the magnetic field in the surroundings of the star is strong enough, polarization vectors (red segments) rotate adiabatically to remain orthogonal to the the external magnetic field lines. Thus, a strong polarization signal coming from the star should be observed.
 \label{Curv}}
\end{figure}

Consider  a light source on the surface of the star. It will emit light along $\bm k$ towards the observer, with a linear polarization orthogonal to $\bm B$ in that point. As the light travels away from the star surface, the direction of  $\bm B$ will effectively change, albeit slowly. However, the finite $\Delta n$ keeps the linear polarization orthogonal to the changing $\bm B$ --- the polarization vector adiabatically follows the variation of the external $\bm B$ along the light path.  Indeed, when sufficiently far from the star, the $\bm B$ field vectors are tangent to a surface with smaller curvature than that at the star's surface (where $R$ is only $\approx 10$~Km) and turn out to be more parallel to each other  than they were on its surface. As a simplified picture, assume for the moment that the magnetic field lines are the tangent vectors along meridians from the north to the south magnetic poles of a  sphere, see Fig.~\ref{Curv}. This approximation is used here for the sake of illustration only and will not be pursued in the rest of the paper

The image of the star which is actually observed is formed in the far region, where all $\bm B$ vectors are almost parallel to each other, and all polarization vectors, being orthogonal point by point to $\bm B$, also result to be almost parallel to each other, differently from how they were on the star's surface.
In the simplified model of magnetic field lines described above, at a latitude of $\theta=\pi/4$ on the sphere, the angle $\psi$ between two magnetic field vectors (tangent to the meridians) taken at some arc distance  $\ell$ with respect to each other, changes with the radius $R$ of the sphere as
\be
(\cos\psi)_R=\frac{1}{2}\left(1+\cos\frac{\ell}{R}\right)
\ee
On the star's surface, at $R=R_{_{\rm NS}}=10$~Km, if we take $\ell=\pi R_{_{\rm NS}}$ we get $\psi=90^\circ$ whereas, on a sphere of radius $R>R_{_{\rm NS}}$, $\psi$ will be smaller, keeping $\ell$ fixed to the same value\footnote{If we take $\ell\simeq \pi R_{_{\rm NS}}$ (with $R_{_{\rm NS}}=10$~Km) the angle between two $\bm B$ vectors at $\theta=\pi/4$, on a sphere with radius $R_{_{\rm NS}}$, is $\psi= 90^\circ$ whereas on a sphere of $R=50$~Km is $\psi\simeq 24^\circ$ and $\psi\simeq12^\circ$ at $100$~Km. Since polarizations follow the magnetic field vectors, as discussed above, a very significant enhancement of the polarization effect is expected to occur even on a length scale of 50~Km.}. 
In light of this an enhanced net polarization signal is predicted. 

The enhancement of the polarization effect due to QED is expected to be so effective that the visible star surface should appear as a superposition of sources all emitting with the same parallel polarizations. In principle a 100\% linear polarization should be measured. 

Of course what is finally observed depends on the degree of polarization of light produced on the surface of the star itself, {\it i.e.} on how well each light source on that surface  is indeed a perfect polarized light emitter. If we assume that this is the case, electrodynamics would suggest a way stronger degree of polarization than what reported. On the other hand, if this were not the case, observing a degree of polarization of the 16\% could also mean that we are still observing the maximum polarization attainable by the QED vacuum effect, but not being sure what is the expected degree of polarization, no strong claim is possible for the first time measurement of the vacuum birefringence predicted by QED.

\begin{figure}[htb!]
\centering
   \includegraphics[width=9truecm]{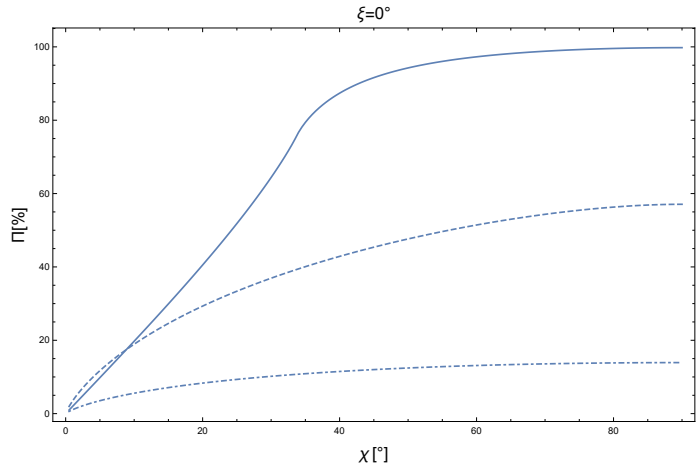}
 \caption{\small $\chi$ is the angle between the rotation axis of the star and the observation line. $\xi$ is the angle between the rotation and magnetic axes.  $\Pi=\sqrt{Q^2+U^2}/I$ is the polarization degree as a function of the two Stokes parameters $Q$ and $U$, whereas $I$ is the total intensity. An average over the period of rotation is done.  
 Dashed lines correspond to the case of no QED effects with a meridian magnetic field. Dot-dashed lines are for the case of no QED effects and a {\it dipole magnetic field}. Solid lines are obtained including the vacuum birefringence effect (with a dipole magnetic field). The results agree almost everywhere, within experimental errors, with those given in~\cite{Mignani:2016fwz}. However, as discussed in the text, our interpretation of this result remains different.  Little variation is found at higher values of $\xi \lesssim 10^\circ$.
 \label{fdue}}
\end{figure}

Even if we assume that  every single point of the star emits polarized light, then  a degree of polarization of $16\%$ may be reached in the absence of QED effects, see Fig~\ref{fdue}. We underscore that the calculations producing Fig.~2 are done both with a simplified magnetic field (meridian lines) and with a realistic {\it dipole field}.   

The results found reasonably agree with those in~\cite{Mignani:2016fwz} (almost everywhere within experimental uncertainties), even though our interpretation of the comparison with  data is rather different from theirs, as we will further explain in the next section.  

It is worth observing that the level of agreement reached  shows how the few effects we neglected are subleading.  For example, the light bending effect studied in Fig.~4 of~\cite{Pavlov:1999cw} --- confront the curves with different $g_r=\sqrt{1-R_S/R}$ in that paper ($R_S$ is the  Schwartzshield radius) --- are indeed known to be quite small, expecially for visible light. 
It is worth observing that, in the case of visible light, relativistic bending increases the polarization by $\sim 5 \div 10 \%$, if QED vacuum birefringence is present \cite{Heyl:GR}. Conversely, if the QED effect is off, relativistic effects on the polarization are $\sim 2 \%$ \cite{Pavlov:1999cw}.

The reason why the degree of polarization $\Pi$ is so high in presence of QED effects (using a dipole magnetic field), can be briefly summarized as follows.
We compute the  approximate formula for $|E_\parallel(z)|^2$, which measures the increase in the polarization component not initially present on the star surface.
\be
|E_\parallel(z)|^2\simeq |E_{\perp}(R_{_{\rm NS}})|^2\, \left|\, \int_{R_{_{_{\rm NS}}}}^z ds\, \phi^\prime(s)\, e^{i\int^s\, dx\, (\lambda_2(x)-\lambda_1(x))}\,\right|^2
\label{soluzioneapp}
\ee
The angle $\phi$, identifying the magnetic field with respect to direction of the radiation direction $\bm k$
\be
\hat{\bm B} =\hat{\bm k} \cos\theta +\hat{\bm x}\,(\sin\theta \cos\phi)+\hat{\bm  y}\, (\sin\theta\sin\phi)
\label{BfieldPolar}
\ee
varies with the distance $z$ from the star surface.  A number of photons have changed their polarization  along the way because  the magnetic field changes not in a perfectly adiabatic way: non-adiabatic transitions from a polarization mode to the other are possible.  
However, from~\eqref{soluzioneapp}, we cannot expect any relevant increase of the mode $E_\perp$  if we have a rapidly oscillating phase in the integrand --- $\phi^\prime(s)$ is almost constant along the path.

This result is consistent with the Zener inequality~\cite{Zener:1932ws} (also known as Landau-Zener theorem) according to which the probability that the polarization switches, in presence of vacuum birefringence, would be
\be
P\leq\left|{\int_{R_{_{NS}}}^{\infty}\varepsilon(z)\, e^{-i \omega\, \Delta n\, z}\,dz \over \int_{R_{_{NS}}}^{\infty}\varepsilon(z) \, dz}\right|^2
\ee
where $\varepsilon(z)$ is some smooth function which is significantly  different from zero only in the transition region and we replace 
\be
\Delta n=\lambda_1-\lambda_2=\frac{(q+m)}{2}
\ee
and 
\be
\label{qetc}
q=7\delta \quad\quad m=-4\delta\quad\quad \delta=\frac{\alpha}{45\pi} \left(\frac{B}{B_{_{\rm QED}}}\right)^2 
\ee
where the critical  magnetic field is given by 
\be
B_{_{\rm QED}}=\frac{m^2}{e}=4.4\times 10^{13}~{\mathrm G}
\ee
The derivation of~\eqref{soluzioneapp} is based on the Euler-Heisenberg Lagrangian. 
See also~\cite{Baier:1967zzc,altrecons,Shakeri:2017knk,Adler:1971wn,Fernandez:2011aa}.

A numerical evaluation of $|E_\perp(z)|^2$ in~\eqref{soluzioneapp}  is  done using $B=10^{13}$~G~\cite{Kerkwijk:2008} and obtaining the $\phi^\prime(s)$ function from the defining equations of the dipolar magnetic field of the star as a function of the distance from its surface. The derivative of the phase factor $S(z)$ in $e^{iS(z)}$ is extremely large $\sim 10^6$ when $z$ is of the order of the radius of the star and rapidly decreases at about 50 Km from the star center.  We take $\omega \approx 1$~eV since measurements in~\cite{Mignani:2016fwz} are done for visible light. Thus we find that within these distances, the rapidly oscillating function $e^{iS(z)}$ makes $|E_\perp(z)|^2$ negligible: if a polarization mode is not initially present on the surface 
of the star it will not be produced along the distance the light travels in getting far out from the star surface. In principle, the variation of $\bm B$ along the light path could  
have been responsible for non-adiabatic transitions between polarization modes, but in practice is not. 
As long as $B\sim B_{_{\rm QED}}$ there are for sure no appreciable non-adiabatic transitions, and even for smaller values of $B$, polarizations will tend to follow adiabatically the variation of $\bm B$. 

One can therefore conclude, on general grounds, that the surroundings of the star ($\sim 100$~Km from its surface), where the QED birefringence is significant, are extremely effective at the polarization enhancement phenomenon described. If every point on the star surface is to be regarded as a perfect linearly polarized light source, then a 
$\approx 100\%$ polarization degree should be observed in the most favorable observation condition.

A complete analysis requires taking into account light bending due to the curved spacetime, non-uniform surface temperature distributions and more complicated magnetic fields, among others~\cite{Gonzalez:2016}. Even if our calculations, briefly summarized in Fig.~2, are capturing the essential features of the phenomenon, in the next section we will  compare the theoretical models, including {\it all} the mentioned effects, with the experimental data reported in~\cite{Mignani:2016fwz}.

{\bf \emph{The measured polarization  compared to theoretical models.}} The conclusions reached  are consistent with a standard statistical analysis carried on  the experimental result presented in~\cite{Mignani:2016fwz} when compared to the same theoretical models chosen in that work. The very fact that data agree at the $1\div 2~\sigma$ level with those models not including QED birefringence, is immediately evident from Fig.~5 in~\cite{Mignani:2016fwz}.   However, in what follows, we want to approach quantitatively this analysis, relying only on the ``Isotropic Blackbody" model as presented in~\cite{Mignani:2016fwz} and further discussed in ~\cite{Gonzalez:2016}. 
 
The experimental result is a ``3-$\sigma$" one,  meaning that, at $\approx 99\%$ C.L.,  the polarization degree is larger than zero. {\it This confidence level is by no means the degree of belief in the existence of vacuum birefringence}. 

Indeed, to ascertain the confidence in the vacuum birefringence hypothesis, it is necessary to compare it against the null one. In other words we  must compare how likely it is that  data come from a theory with and without vacuum birefringence. This is classically done through the calculation of a Bayes factor. We have therefore
\bea
&& H_0 = \text{100\% polarization at the star surface {\it and} no vacuum birefringence}\notag\\
&& H_1 = \text{100\% polarization at the star surface {\it and} vacuum birefringence}
\eea
The angle $\chi$ between the rotation axis and the Line Of Sight (LOS), the direction from the observer to the star, is not known exactly. Before the experimental measurement, we have a joint prior probability density function for $H_i$ and $\chi$, given by $f_0(H_i, \chi) = P_0(H_i)\cdot f_0(\chi)$ --- we indicate with $f$ a probability density and with $P$ a probability.  The factorization can be made assuming, as is perfectly reasonable, that the stellar theory $H_i$ is independent from the contingent angle $\chi$ (a random orientation in space). 

After the experimental measurements, the probabilities are updated using the conditional probability theorem and taking the ratios of the alternative hypotheses
\be
\frac{ f (H_0,\chi | \text{data} ) }{ f(H_1,\chi | \text{data})} = \frac{f(\text{data} | H_0,\chi)}{f(\text{data} | H_1,\chi)}  \frac{P_0 (H_0) f_0(\chi)}{P_0 (H_1) f_0(\chi)}
\ee
where $f(H_i,\chi | \text{data})$ are the posterior distributions. We marginalize on the angle $\chi$
\be
\frac{P(H_0 | \text{data})}{P(H_1 | \text{data})} = \frac{\int f(\text{data} | H_0,\chi) f_0(\chi) d\chi }{\int f(\text{data} | H_1,\chi) f_0(\chi)  d\chi} \times \frac{P_0(H_0)}{P_0(H_1)}=L \frac{P_0(H_0)}{P_0(H_1)}
\label{eqbayes}
\ee
The ratio 
$L$  is the Bayes factor, or likelihood ratio. It  tells how much the probabilities of the alternative hypotheses being true change based on the experiment. In this case a Bayes factor $L\gg1$ indicates that the data favors absence of vacuum birefringence, while a value $L\ll1$ would favor its presence. An experiment claiming to be proof of vacuum birefringence should, at the very least, have a Bayes factor significantly {\it smaller} than $1$.

We add that there is a further unknown parameter $\xi$: the angle between the magnetic field axis and the rotation axis. Mutatis mutandis, the above argument simply changes to include a joint probability $f_0(\chi,\xi)$, and an integration over all $\xi$.  Measurement of the X-ray pulsed fraction for RX J1856.5$-$3754, constrains the values of $\chi$ and $\xi$. This constraint is synthesized by saying that the star has a small angle $\xi \lesssim 6^\degree $ while  $\chi \approx 20^\degree  \div  45^\degree$, which may arguably be larger\cite{Ho:2007gs}.

Based on this  we construct prior probability distributions $f_0(\chi, \xi)$. Ideally, we would use posterior distributions estimated from previous theoretical analysis of the X-ray pulsed fraction, but only upper and lower limits are provided. We will therefore conduct our calculation with three different priors, and then show that the qualitative result is largely independent of these. In the first case $A$ we use 
\be
	f_{0,A}(\chi,\xi) \propto \sin \chi \exp\left( -\frac{\xi}{6^\degree}\right)
\ee
This is equivalent to take $\chi$ and $\xi$ to be independent, $\cos \chi$	to be uniform in the interval $[0,1]$ and $\xi$ to be exponentially distributed with a mean of $6^\degree$. This  captures the result of the analysis in~\cite{Ho:2007gs} for $\xi$ (on which~\cite{Mignani:2016fwz} relies), while leaving us ignorant on $\chi$. In the second case $B$, we use 
\be
f_{0,B}(\chi,\xi) \propto \exp\left[ - \frac{(\chi - 32^\degree)^2}{2 (12^\degree)^2}\right] \exp\left( -\frac{\xi}{6^\degree}\right)
\ee
where $\xi$ is distributed as before. We take $\chi$ to be normally distributed with a mean of $32^\degree$ and standard deviation of $12^\degree$ which corresponds to taking the interval quoted by~\cite{Ho:2007gs} to be a 68\% confidence interval. Finally we consider the less  motivated case, case $C$, which corresponds to a flat prior in the cosine of both variables. 

Given the hypothesis $H_i$ and the angles $\chi$ and $\xi$, there is an expected theoretical polarization degree $\Pi(H_i,\chi,\xi)$. The distributions $f(\text{data} | H_0,\chi,\xi)$, which appear in the computation of $L$ are taken to be
\be
f(\text{data} | H_i,\chi,\xi) \propto \frac{1}{\sqrt{2\pi \sigma^2}} \exp\left[ - \frac{(\bar \Pi-\Pi(H_i,\chi,\xi))^2 }{ 2 \sigma^2}\right]
\ee
where $\bar \Pi = 16.4\%$ is the experimentally measured polarization degree and $\sigma=5.2\%$ the experimental error \cite{Mignani:2016fwz}. 

The likelihood ratio $L$ is evaluated numerically. The results are collected in Table~\ref{tableLikelihood}.  For both  $A,B$ priors considered, we find $L > 1$. We conclude that the data, when compared to models, favor the {\it absence} of vacuum birefringence. This is in strong contrast with the qualitative claim done in~\cite{Mignani:2016fwz}.  In the $C$ case, the less motivated one, there is no significant discrimination between the two hypotheses. 
Data are taken from Figs.~3 and~5 in~\cite{Mignani:2016fwz}. This result is not unexpected: when comparing two hypotheses, the one which would place a more stringent constraint on a unknown parameter is disfavored, unless there is a strong prior belief on the value of the unknown parameter. In other words, while $H_0$ gives a definite prediction on $\bar \Pi$, $H_1$ predicts nearly every possible value (for example Fig. Fig~\ref{fdue}).

\begin{figure}[htb!]
	\centering
	\includegraphics[width=8cm]{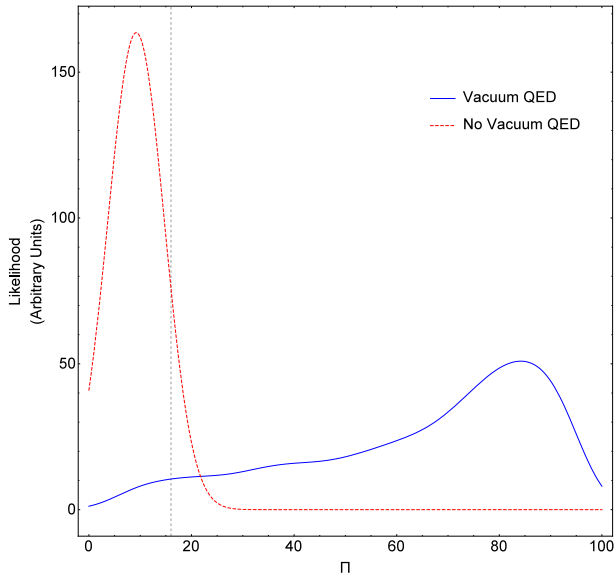}
	\caption{\small The likelihood of the polarization degree in both hypotheses including the integration over the unknown angles $\chi$ and $\xi$. Prior probabilities of case $B$ have been used. The vertical dashed line is the observed value. In order for experimental data to favor the existence of vacuum birefringence,  values of the polarization degree  larger than $\approx 30\%$ must be measured.
		\label{FigLikelihoods}}
\end{figure}

The same methods can be used to estimate how much polarization degree must be observed, assuming fixed experimental error $\sigma$, in order for the data to favor the presence of vacuum birefringence. We estimate a polarization degree of $\bar \Pi\approx 22\%$ must be observed for $L \approx 1$ and $\bar \Pi\approx 29\%$ in order for $L \approx 0.01$, a more solid result; see Fig~\ref{FigLikelihoods}. Furthermore, we estimate that if the experimental error were reduced, in the future, with $\sigma =3\%$ a polarization degree of $\Pi\approx 23\%$ would be needed to be measured so that $L\approx 0.01$.

Using the above method, one can compare different light and emission propagation models by calculating likelihood ratios. For example, one may consider the hypothesis in which vacuum birefringence exists but the light is not $100\%$ polarized at the star's surface, and compare this hypothesis with $H_0$.
\begin{table}[htb!]
\begin{center}
\begin{tabular}{c | c | c | c}
\hline\hline
	& Case $A$ & Case $B$ & Case $C$ \\
	\hline
	$L$ defined in~\eqref{eqbayes}  & $28.7$ & $7.46$ & $0.65$ \\
	\hline\hline
\end{tabular}
\end{center}
\caption{Likelihood ratios in the three cases described above. In both the the absence of vacuum birefringience effects is favored. This table is incompatible with any {\it strong} claim in either direction.}
\label{tableLikelihood}
\end{table}

Results displayed in Fig.~\ref{FigLikelihoods} are obtained assuming a specific model whose validity is uncertain. This statistical analysis could provide even more adverse results if all possible sources of uncertainty were considered.

{\bf \emph{Conclusions.}}  The effects of QED vacuum birefringence have never been experimentally observed, but searched at length, over the years, in laboratory  experiment  as PVLAS (see e.g.~\cite{Bregant:2008yb} and references therein). The possibility that some stars, known as Magnetars, could have magnetic fields as large as $10^{13}$~G opens certainly an interesting perspective for a different way of studying this phenomenon. However we conclude that the  claim of the first observation of a QED vacuum birefringence effect,  raised in~\cite{Mignani:2016fwz}, cannot be considered as conclusive and this is  not (only) for the reason that the polarization signal is, for the moment, only  a $\approx 3\sigma$ effect. 

We conclude that only rather high degrees of linear polarization ($\gtrsim 30\%$), see Fig~\ref{FigLikelihoods}, would be the indisputable  footprints of QED birefringence effects,  confirming that the star surface emits polarized light, as claimed by several authors, and that the star surroundings, being pervaded by a magnetic field $B\sim B_{_{\rm QED}}$,  indeed force the light polarization vectors to adiabatically follow the magnetic field orientation thus becoming almost parallel as in Fig.~\ref{Curv}. 

As  from Fig.~\ref{FigLikelihoods}, the hypothesis with no birefringence effect is even more significant than the one including the effect. This conclusion, reached on the basis of a standard statistical analysis, is in good agreement with what was proposed in the first version of our paper: measuring a degree of polarization larger than $\approx 40\%$ would give a different reliability to  claims of `strong evidence of vacuum birefringence effects'.    
Data and models discussed in~\cite{Mignani:2016fwz} (or in the note~\cite{Turolla:2017tqt}) have been used exclusively. 
That said, it would be of extreme interest to confirm and bring to a better statistical significance the results discussed in~\cite{Mignani:2016fwz}.

For any axion-like particle contribution to have a role in changing the results of our analysis, which is limited to light in the visible spectrum $0.1<\omega<2$~eV (the inclusion of photons to axion-like particles conversions is that of shifting $q$ in~\eqref{qetc} by $q\to q+B^2\,f(G/m_a,\omega/m_a)$), one should have photon-axion couplings of the order of $G \approx 10^{-7}$~GeV$^{-1}$ for all  $m_a$ values which however are found in the region already ruled out in the exclusion plot reported in~\cite{pdg}.

\section*{Acknowledgements}
We thank George Pavlov for  comments on our manuscript. We thank Daniele Del~Re for very useful discussions on the statistical analysis and Claudio Gatti for  interesting comments. 

\bibliographystyle{unsrt}

\end{document}